\documentclass[dvips]{article}
\usepackage{amsmath}    % math package
\usepackage{amsfonts}
\usepackage{amsthm}
\usepackage{graphicx}   % extended graphics package 
\usepackage{array}      % extended styles for tables
\usepackage{hhline}     % extended styles for tables
\usepackage{pslatex}    % so that we've got a clean pdf file afterwards
\usepackage{natbib}
\usepackage{bbm}
\usepackage{icrctc07}

\title{Large scale magnetic field of the Milky Way from WMAP3 data}
\shorttitle{Magnetic Field of the Milky Way}
\authors{R. Jansson$^{1}$,  G. R. Farrar$^{1}$, A. H. Waelkens$^{2}$, T. A. Ensslin$^{2}$}
\shortauthors{R. Jansson and et al}
\afiliations{$^1$Center for Cosmology and Particle Physics,
and Department of Physics\\ New York University, NY, NY 10003,USA\\$^2$Max-Planck-Institute f\"ur Astrophysik, Karl Schwarzschild-Str. 1, 85741 Garching, Germany}
\email{rj486@nyu.edu}

\abstract{We report on initial results from a project to constrain the large-scale and turbulent magnetic fields of the Milky Way galaxy, which eventually will incorporate all of the relevant observational data.  In this paper we fit popular large scale magnetic field models to WMAP3 polarization maps. We find that the polarization data can constrain certain model parameters but does not uniquely determine the best-fit parameters. We also find that the polarization data alone cannot distinguish between model symmetries, e.g., the existence of field reversals. We show how future UHECR data can break this degeneracy.}

\begin{document}
\maketitle
%Begin the section.

\section{Introduction}

A variety of observational methods has been used to constrain the magnetic field of the Milky Way, e.g., synchrotron emission, Faraday rotation, Zeeman splitting and polarization of starlight. In a project currently underway also incuding M. Haverkorn and A. Lazarian we make use of all relevant data available to study the Galactic magnetic field (GMF), at large and small scales. The reason for combining different types of data goes beyond having a larger number of data points when performing parameter estimation. Different types of data constrain different aspects of the magnetic field. For instance, synchrotron emission probes the integrated magnetic component perpendicular to the line-of-sight while Faraday rotation and starlight polarization probe the parallel component. Zeeman splitting allow \emph{in situ} measurements of the magnetic field.

In this paper we present first results, using polarization data from the Wilkinson Microwave Anisotropy Probe (WMAP) 3rd year data release \cite{wmap_pol:2006}. A previous analysis of fitting the WMAP3 data to a large scale magnetic field model was done in \cite{wmap_pol:2006}. There the WMAP team fitted the polarization angle of synchrotron emission as predicted by an axisymmetric magnetic field to the measured polarization angle ($\gamma=\frac{1}{2}\tan^{-1}(U/Q)$), after masking out 25.7\% of the sky (essentially the Galactic disk and the North Galactic Spur).

\section{Galactic magnetic field models}

When describing models of the regular part of Galactic magnetic fields it is natural to use cylindrical coordinates $(r,\theta,z)$. 
Typically, the disk field is then parametrized by
\begin{equation}
B_r = B(r,\theta) \cos p, \hspace{0.7cm}  B_\theta = B(r,\theta) \sin p,
\end{equation}
where $p$ is the pitch angle. There has been some confusion in notation in the literature, and we will use the following conventions. A field that obeys  $B(r,\theta)=B(r,\theta+\alpha)$ for any $\alpha$ is called \emph{axisymmetric}, if $B(r,\theta)=-B(r,\theta+\pi)$ it is called \emph{bisymmetric}, and if $B(r,\theta)=B(r,\theta+\pi)$ we propose to call it \emph{disymmetric}.
Another distinction that can be made is a model's symmetry properties under $z\rightarrow -z$, i.e., reflection at the disk plane. We call a field symmetric with respect to the Galactic plane if $\textbf{B}(r,\theta,-z)=\textbf{B}(r,\theta,z)$, and antisymmetric if $\textbf{B}(r,\theta,-z)=-\textbf{B}(r,\theta,z)$. This notation agrees with, e.g., \cite{Tinyakov:2001,Harari:1999}, but conflicts with, e.g.,  \cite{Stanev:1997}.

A standard way to parametrize a bisymmetric spiral field \cite{Han:1994} is
\begin{equation}\label{Han_B}
 B(r,\theta) = b(r) \cos (\theta-\beta \ln \frac{r}{r_0}),
\end{equation}
where $\theta$ is defined to increase clockwise around the Galactic center, and $\theta = 0$ points to the Sun. The pitch angle $p$ is positive if the clockwise tangent to the spiral is outside a circle of radius $r$ centered on the Galactic center, and $\beta = 1/\tan p$. At the point $(r_0, \theta=0^\circ)$, the field reaches the first maximum in the direction $l=180^\circ$ outside the solar circle. 

We choose $b(r) = b_0$ for $r<r_c$, and  $b(r) = b_0\frac{r_c}{r}$ for $r>r_c$, and define the field for $r\leq20\,$kpc.

\section{Using WMAP3 polarization data}

Relativistic electrons accelerated by a magnetic field will radiate synchrotron radiation \cite{Rybicki:1986}. For a power law distribution of cosmic ray electrons, $n(E)dE \sim E^{-s}dE$, the synchrotron emissivity is 
\begin{equation}
j_\nu\propto B_\perp^{\frac{1+s}{2}}\nu^{\frac{1-s}{2}}.
\end{equation}
The emission has a large degree of linear polarization.

WMAP is an all-sky survey measuring the Stokes $I,Q,U$ parameters in five frequency bands in the 23-94 GHz range. The Stokes $I$ parameter is proportional to the total (polarized and unpolarized) emission, and the $Q$ and $U$ parameters describe the polarized part of the emission. In the K band (23 GHz) the measured polarized radiation is dominantly Galactic synchrotron emission.

We obtain the best fit by minimizing $\chi^2$ for $Q$ and $U$. The resolution in the K band is $\sim 1^\circ$, but to better probe the large scale regular field we smooth the data to $\sim 4^\circ$. For each $4^\circ$ pixel we calculate the variance of the $1^\circ$ pixels within a radius $\rho$ from the center position of the larger pixel. This way we obtain $\sigma_Q$ and $\sigma_U$ for the $4^\circ$ maps. Regions with large turbulence and irregular structures have a larger variance and thus are deweighted in the $\chi^2$ fit. We thus avoid masking out the Galactic disk, which is essential to be able to probe the magnetic structure in the plane of the Galaxy. For a given Galactic magnetic field and electron distribution (using the Hammurabi code \cite{Waelkens:2008}) we calculate the polarized synchrotron emission $Q$ and $U$. We then measure the goodness-of-fit by 
\begin{equation}
\chi^2_Q = \sum_{i=1}^{pixels}\frac{(Q^{model}_i-Q^{data}_i)^2}{\sigma^2_{Q,i}},
\end{equation}
and similarly for $U$, with $\chi_{QU}^2=\chi^2_Q+\chi^2_U$ the quantity that is minimized.

\begin{figure*} [h]
\begin{center}
% \noindent \fbox{\hbox{\vbox{\hsize=130mm \hfill \vspace{50mm}}}}
\includegraphics [width=1\textwidth]{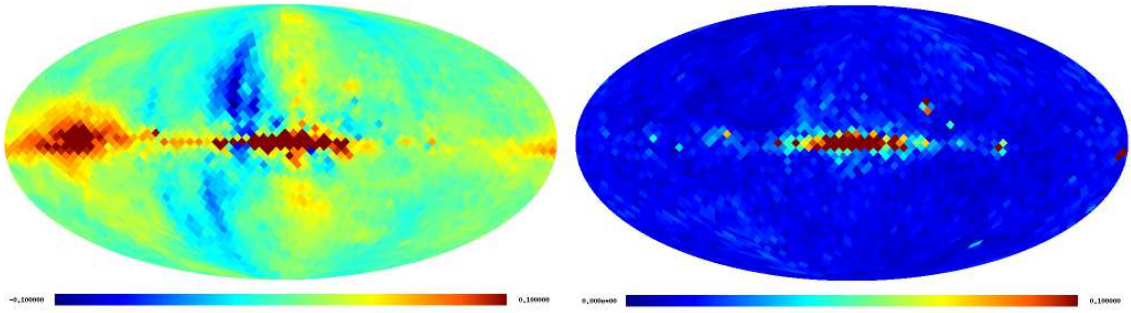}
\end{center}
\caption{\emph{Left:} Stokes Q parameter in mK from WMAP3 smoothed to $4^\circ$ resolution. \emph{Right:}   $\sigma_Q$ in mK calculated with $\rho = 2^\circ$.}\label{sigma_Q}
\end{figure*}

In figure \ref{sigma_Q} a smoothed map of the WMAP3 $Q$ parameter is shown, along with map of $\sigma_Q$ obtained using $\rho = 2^\circ$.

\section{Results}

\begin{figure*} [h]
\begin{center}
% \noindent \fbox{\hbox{\vbox{\hsize=130mm \hfill \vspace{50mm}}}}
\includegraphics [width=1\textwidth]{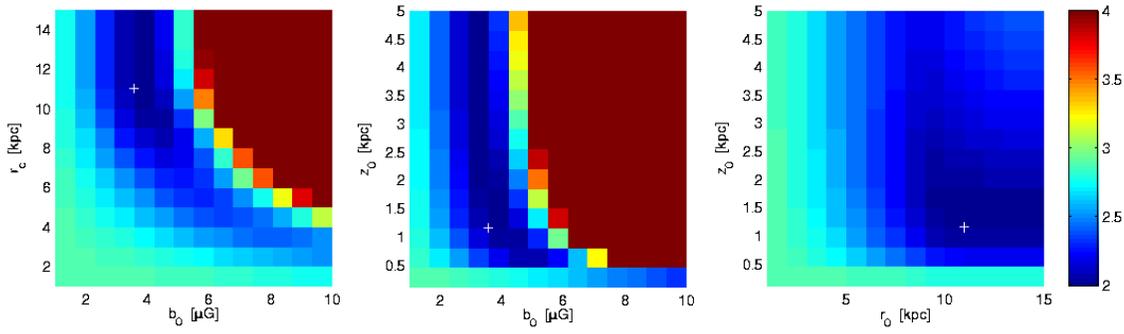}
\end{center}
\caption{$\chi^2$ surface plots for bisymmetric field model. White crosses mark best-fit values.}\label{chi_bss}
\end{figure*}

We perform a parameter search to minimize $\chi^2_{QU}$ for the parameters $b_0$, $r_c$ and $z_0$ for a bisymmetric spiral field that is symmetric in $z$. Following Stanev \cite{Stanev:1997}, we put $p=-10^\circ$ and $r_0=10.55\,$kpc. From figure \ref{chi_bss} it is evident that the data strongly exclude large regions of parameter space. However, no unique best-fit parameters are found, as a number of combinations of the search parameters yield a $\chi^2$ very close to the minimum. 

To investigate the effect of model symmetries, we perform the same analysis on a bisymmetric field that is antisymmetric in $z$, and two disymmetric field models (symmetric/antisymmetric in $z$). We obtain the disymmetric model from equation \ref{Han_B} by squaring the cosine. We find that the minimized $\chi^2$ is almost identical for all four models.   

\begin{figure*} [h]
\begin{center}
% \noindent \fbox{\hbox{\vbox{\hsize=130mm \hfill \vspace{50mm}}}}
\includegraphics [width=1\textwidth]{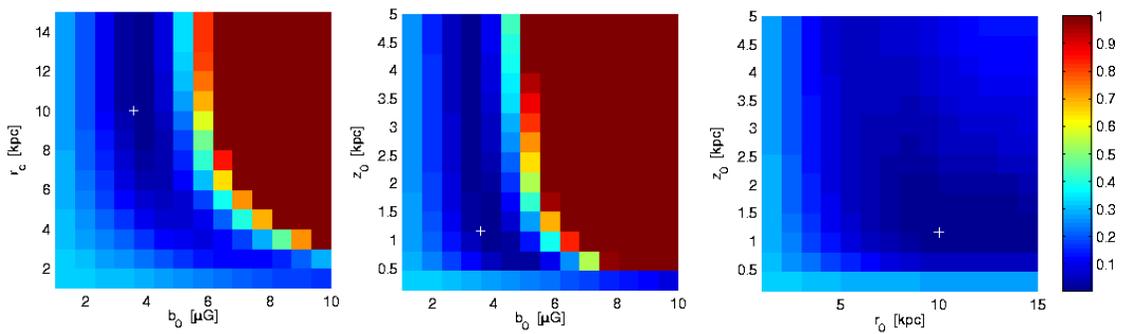}
\end{center}
\caption{$\chi^2$ surface plots for bisymmetric field model fitted to mock data from the same bisymmetric GMF model. $\chi^2$ values only shown for relative comparison.}\label{chi_fake_bss}
\end{figure*}

To examine further the limitations of using polarization maps to constrain GMF models we produce mock $Q$ and $U$ maps with Hammurabi using a $z$-symmetric bisymmetric field model with the parameter values $b_0=3.5\,\mu G$, $r_c=10\,$kpc, $z_0=1.15\,$kpc, $p=-10^\circ$ and $r_0=10.55\,$kpc. Using these maps as ``data'', and with uniform $\sigma$, we do a parameter search for the same bisymmetric field used to obtain the mock data. The result is shown in figure \ref{chi_fake_bss}. We note that the correct best-fit values are recovered, but that the insensitivity to the model parameters remain. If a $z$-antisymmetric disymmetric model is fitted to the above mock data we find the $\chi^2$ surfaces to be visually indistinguishable to those in figure \ref{chi_fake_bss}, and we can conclude that even with ideal data, polarization maps alone cannot distinguish between the various model symmetries.
 
\begin{figure} [h]
\begin{center}
% \noindent \fbox{\hbox{\vbox{\hsize=130mm \hfill \vspace{50mm}}}}
\includegraphics [width=.5\textwidth]{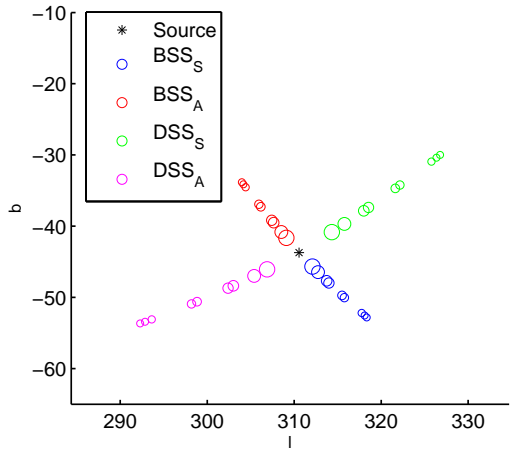}
\end{center}
\caption{A hypothetical UHECR multiplet source along with the UHECR locations as predicted from bisymmetric and disymmetric spiral GMF models. Subscripts refer to symmetry/antisymmetry under reflection in $z$.  UHECR energies selected in the $10^{19-20}$ EeV range, with the size of the circles proportional to the event energies.}\label{multiplet}
\end{figure}

\section{UHECR multiplets}

A powerful way to break this degeneracy between model symmetries would be to use future UHECR multiplets. UHECR deflection is unique among observables in being independent of electron density and very weakly dependent on random fields because the correlation length of the random fields are much smaller than the Larmor radius \cite{TTMWrand}. As the deflection angle of a UHECR is inversely proportional to its energy and proportional to the transverse magnetic field, the arrival directions of an UHECR multiplet will roughly form a string on the sky; the highest energy cosmic ray closest to the direction of the source, and the lower energy cosmic rays further removed according to their energy. From their energies and angular position on the sky their common source can be estimated. A hypothetical multiplet placed in the southern hemisphere is shown in figure \ref{multiplet}, together with the arrival directions of the multiplet as predicted for the best fit spiral field parameters with the four different GMF symmetries. It is clear that if a number of such multiplets would be discovered in future experiments they would be very useful in breaking the degeneracy of the model symmetries.

\section{Summary}

We have reported on a project that will eventually make use of all relevant data to constrain models of the Galactic magnetic field. In this paper we outline our analysis using the WMAP3 polarization data. We improve on the previous analysis \cite{wmap_pol:2006} of the same data by including the Galactic plane, and allow for weighting individual pixels based on the variance in the data. 
 
We find that polarization maps alone can exclude large regions of parameter space for GMF models common in the literature, but fail to give a unique set of best-fit parameters. Furthermore, we find the data cannot distinguish between fundamental model symmetries, such as the existence of field reversals between spiral arms. We comment on the usefulness of using UHECR multiplets to break this degeneracy, if such multiplets were to be found.

In future work random fields will be introduced and the synchrotron component of $I$, in addition to $Q$ and $U$ will be used in the fit, as well as Faraday Rotation Measures and starlight polarization data. More details will be given in a forthcoming paper.

\section{Acknowledgements}
This research has been supported in part by NSF-PHY-0401232.

%This is the reference to .bib file (Whitout .bib!)
\bibliography{icrc1271}
%This in the bibtex style, is ok.
\bibliographystyle{plain}

\end{document}